 \documentclass[aoas,preprint]{imsart}
\usepackage{amsthm,amsmath,chicago}
\RequirePackage{graphicx,color,verbatim,multirow,pgf,pgfarrows,pgfnodes}

\usepackage[font=small,format=plain,labelfont=bf,up,textfont=normal,up,justification=justified,singlelinecheck=false]{caption}
\usepackage{array}
\newcolumntype{$}{>{\global\let\currentrowstyle\relax}}
\newcolumntype{^}{>{\currentrowstyle}}

\usepackage{longtable}
\usepackage{tabu}

\startlocaldefs
\endlocaldefs

\begin{document}

\begin{frontmatter}

\title{Estimating HIV Epidemics for Sub-National Areas}
\runtitle{Estimating HIV Epidemics for Sub-National Areas}


\author{\fnms{Le} \snm{Bao$^1$}\ead[label=e1]{lebao@psu.edu}}
\and
\author{\fnms{Xiaoyue} \snm{Niu$^1$}}
\and
\author{\fnms{Mary} \snm{Mahy$^2$}}
\and
\author{\fnms{Peter D.} \snm{Ghys$^2$}}
\affiliation{Department of Statistics, Penn State University, University Park, PA, USA$^1$\\
UNAIDS, Strategic Information and Evaluation Department, Geneva, Switzerland$^2$\\
\printead{e1}}

\runauthor{L. Bao et. al.}

\begin{abstract}
As the global HIV pandemic enters its fourth decade, increasing numbers of surveillance sites have been established which allows countries to look into the epidemics at a finer scale, e.g. at sub-national levels. Currently, the epidemic models have been applied independently to the sub-national areas within countries. However, the availability and quality of the data vary widely, which leads to biased and unreliable estimates for areas with very few data. We propose to overcome this issue by introducing the dependence of the parameters across areas in a mixture model. The joint distribution of the parameters in multiple areas can be approximated directly from the results of independent fits without needing to refit the data or unpack the software. As a result, the mixture model has better predictive ability than the independent model as shown in examples of multiple countries in Sub-Saharan Africa.
\end{abstract}


\end{frontmatter}

\section{Introduction}
\label{sect-Introduction}
Since the first case reported in 1981 \cite{CDC1981}, the global HIV epidemic has become one of the greatest threats to human health and development. The number of people living with HIV worldwide continued to grow, reaching 35 million in 2013, about three times more than in 1990. There have been 78 million infected with HIV and over 29 million AIDS-related deaths so far. The response to the HIV epidemic has been mixed, with progress being made to reduce new infection and rapid improvements in survival rates among AIDS patients as antiretroviral therapy has become available in recent years \cite{UNAIDS2014}. 

Countries need to ground their HIV strategies in an understanding of their own epidemics and their national responses. Reliable estimation and prediction of the HIV epidemic can help policy makers and program planners efficiently allocate their resources, plan and manage the intervention, treatment and care programs, evaluate their effort, and raise funds. Therefore, accurate estimation and projection of the epidemic is the foundation of all HIV-related studies \cite{WHO2013}. In addition, countries -- especially large countries and countries with geographically heterogeneous epidemics -- need to understand the variation of the impact within the country by geographic area and population \shortcite{Mahy2014}. Having surveillance data and HIV estimates available by sub-national entities helps allocating limited resources \cite{UNAIDS2013}. 


Most countries (158 countries submitted files in 2014) use Spectrum software to estimate the impact of HIV on their population at the national level \shortcite{Case2014}.  Spectrum relies on data from HIV surveillance systems and the Estimation and Projection Package (EPP) to estimate trends in HIV incidence over time \shortcite{Spectrum2014,Brown2014,Stover2014}. The UNAIDS Reference Group on Estimates, Modelling and Projections has developed, and continues to refine Spectrum/EPP for estimation and short-term prediction of HIV trends \shortcite{Ghys2004,Brown2006,Brown2010,Brown2014}. Due to the paucity of reliable information on the incidence of HIV in most countries, sentinel surveillance systems for HIV were designed to monitor the HIV prevalence trends. Using these surveillance data from 1970 through the current year, EPP reconstructs trends in HIV prevalence, incidence and mortality and makes short term projections.  



As the global HIV pandemic enters its fourth decade, increasing numbers of surveillance sites have been established which allows countries to look into the epidemics at a finer scale, e.g. at sub-national level. Most countries with generalized epidemics (epidemics in which HIV transmission is primarily in the general population through heterosexual sex) have historical HIV surveillance data from women attending antenatal clinics from year 2000 to present.  In addition, all but four generalized epidemic countries have household surveys that measure the HIV prevalence in the national population as well as sub-national estimates to the first sub-national administrative level \shortcite{Marsh2014}.  Currently, the estimation and projection process has been implemented independently for areas in those countries. However, the availability and quality of HIV surveillance data used for these models is variable \shortcite{Calleja2010}. Some key components of data availability and quality include the number of years of data to show trends over time, the representativeness of the data across the country, and the accuracy of those data \shortcite{Lyerla2008}. When the sub-national areas do not have sufficient high quality data the models produce inaccurate results with large uncertainty bounds. 

 
One way to improve accuracy of the estimates is to borrow information from neighboring areas in the case of sub-national estimates. For example many areas have few data points early in the HIV epidemic (between 1980 and 2000). In such situation, if the data from neighboring areas are assumed to have similar trends, the trends from those areas can be used to inform the one in the lack-of-data area. In this paper we introduce an informative prior distribution derived from the mixture model to allow sharing information across datasets. In the mixture model, the epidemiologically implausible trajectories fit in the area with few data would be down-weighted by the area with rich data. The difficulty of utilizing the proposed model is that the underlying epidemic model is described by differential equations, and the computational cost of fitting such a model is high because the model parameters do not have analytic solutions. In this article, we propose an importance sampling method that draws posterior samples of parameters in the new model without needing to refit the data. 


In Section 2 we describe the EPP model used by UNAIDS, the mixture model, the parameter estimation, and the evaluation procedure in detail. In Section 3, we give results for 20 countries in Sub-Saharan Africa, and in Section 4 we offer conclusions and discussion for future work. 



\section{Approach}
\label{sect-Approach}
In this section, we first describe a dynamic model used by EPP for estimating and projecting the HIV epidemic from prevalence data. We then introduce the mixture model that allows information to be shared across areas with similar epidemiological patterns. After that we propose a novel parameter estimation method for the mixture model that avoids additional runs of the dynamic model. Finally we discuss the model evaluation procedure.

\subsection{The EPP Model}
The Estimation and Projection Package (EPP) is based on a simple susceptible-infected (SI) epidemiological model.  The dynamic model is as follows :
\begin{equation}
\left\{\begin{array}{ccc}
\frac{dZ(t)}{dt} & = & E(t) - r(t) \rho(t) Z(t) - \mu(t) Z(t) - a_{50}(t)Z(t) +  M(t)Z(t), \\
\frac{dY(t)}{dt} & = & r(t) \rho(t) Z(t) - \textup{HIVdeath}(t) - a_{50}(t)Y(t) +  M(t)Y(t), \\
\end{array}\right.
\label{eqn:ode}
\end{equation}
The population being modeled is aged between 15 and 49, and the population at time $t$ is divided into two groups: $Z(t)$ is the number of uninfected individuals, $Y(t)$ is the number of infected individuals, $N(t) = Z(t) + Y(t)$ is the total adult population size, $E(t)$ is the number of new adults entering the population, $\rho(t)=Y(t) / N(t)$ is the prevalence rate, and $\textup{HIVdeath}(t)$ is the number of deaths among infected individuals. There is a set of parameters defined by external life-tables such as $\mu(t)$, the non-AIDS mortality rate; $a_{50}(t)$, the rate at which adults exit the model after attaining age 50, and $M(t)$, the rate of net migration into the population. 

The infection rate, $r(t)$, is the expected number of persons infected by one HIV positive person in year $t$ in a wholly susceptible population $Z(t)$, and is assumed to have the following form:
\begin{equation}
\log r(t+1)-\log r(t) = \beta_1 \times (\beta_0 - r(t)) - \beta_2 \rho(t) + \beta_3 \gamma(t),
\label{eqn:rtrend}
\end{equation}
where $\gamma_t = \frac{(\rho(t+1)-\rho(t)) (t-t_1)^+}{\rho(t)}$ implies the tendency of stabilization after year $t_1$. Therefore, Equation (\ref{eqn:rtrend}) imposes some common structure on the changes of $r(t)$ across countries: $r(t)$ declines if $r(t)>\beta_0$, or $\rho(t)$ is too high, or the relative increase of prevalence is too large. Furthermore, we define $t_0$ as the start year of the epidemic at which the prevalence is 0.0025\%, and let $r_0$ be the infection rate at $t_0$.

The data for EPP model estimation mainly come from ANC data which consist of the number of infected women $Y_{it}$ and the number of women tested $N_{it}$, for a given year $t$ and a given clinic $i$. Let $W_{it}=\Phi^{-1}(\frac{Y_{it}+0.5}{N_{it}+1})$, where $\Phi^{-1}(.)$ is the inverse cumulative distribution function of the standard normal distribution. The data and the population prevalence -- $\rho(t)$ -- derived from the dynamic system (\ref{eqn:ode}) are linked through a random effect model:
\begin{eqnarray}
W_{it}&=&\Phi^{-1}(\rho_t)+\beta_4+b_i+\epsilon_{it}, \nonumber\\
b_i&\sim& N(0,\sigma^2), \label{eqn:ranef} \\
\epsilon_{it}&\sim&N(0,\nu_{it}) \nonumber
\end{eqnarray}
where $\beta_4$ is the bias of ANC data with respect to prevalence data from  national population-based household surveys (NPBS) \cite{Bao2012rtrend}, and $b_i$ is the clinic random effect. $\sigma^2$ is assumed to have an inverse-Gamma prior which gets integrated out in the likelihood evaluation. $\nu_{it}$ is a fixed quantity that depends on the clinic data and approximates the binomial variation \cite{Alkema2007}.

With each input set, $\boldsymbol{\theta} = (t_0, t_1, r_0, \beta_0, \beta_1, \beta_2, \beta_3, \beta_4)$, from Equation (\ref{eqn:ode}) and (\ref{eqn:rtrend}), there is a series of output $\rho(t)$, which get evaluated in the likelihood based on Equation (\ref{eqn:ranef}). With a set of informative priors incorporating empirical evidence and expert opinions as in Equations (\ref{eqn:prior}), Bayesian inference of the population prevalence, incidence and mortality (all derived from the dynamic system) can be performed for the EPP model \cite{Bao2012rtrend}. To be specific, the following informative prior distributions are used:
\begin{equation}
\begin{array}{ccl}
t_0 & \sim & \textup{Uniform}[1970, 1990], \\
t_1 & \sim & T_{(2)}(20, 0.458), \\
\log r_0 & \sim & T_{(2)}(0.42, 0.081), \\
\beta_0 & \sim & T_{(2)}(0.46, 0.192), \\
\beta_1 & \sim & T_{(2)}(0.17, 0.091), \\
\beta_2 & \sim & T_{(2)}(-0.68, 0.264), \\
\beta_3 & \sim & T_{(2)}(-0.038, 0.009),\\
\beta_4 & \sim & T_{(2)}(0.14, 0.068).
\end{array}
\label{eqn:prior}
\end{equation}
where $T_{(2)}(\mu_0, \sigma_0^2)$ is a t-distribution with location parameter $\mu_0$, scale parameter $\sigma_0$, and 2 degrees of freedom. The $T_{(2)}$ distribution has infinity variance which ensures a good coverage of all possible parameter values. For simplicity, $(t_0, t_1, r_0, \beta_0, \beta_1, \beta_2, \beta_3, \beta_4)$ are assumed to be independent in the prior distribution. The dynamic model runs at a 0.1-year time step, so $t_0$ and $t_1$ will be rounded into one decimal place when calculating the likelihood.

\subsection{The Mixture Model}
\begin{figure}[!h]
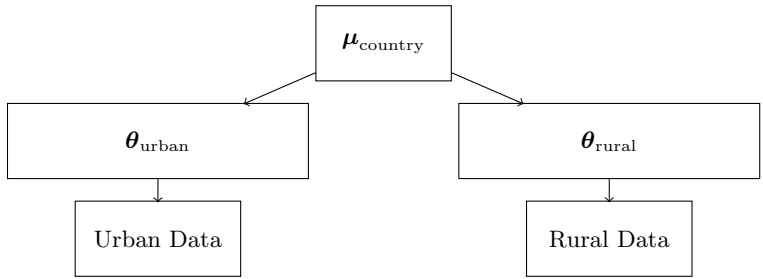

\begin{center}
\begin{pgfpicture}{-2.5cm}{-1.5cm}{2.5cm}{0.5cm}
\pgfnoderect{country}[stroke]{\pgfxy(0,0)}{\pgfpoint{1.8cm}{1cm}}
\pgfnoderect{urban}[stroke]{\pgfxy(-3,-1.3)}{\pgfpoint{4cm}{1cm}}
\pgfnoderect{rural}[stroke]{\pgfxy(3,-1.3)}{\pgfpoint{4cm}{1cm}}
\pgfnoderect{data1}[stroke]{\pgfxy(-3,-2.6)}{\pgfpoint{2.2cm}{1cm}}
\pgfnoderect{data2}[stroke]{\pgfxy(3,-2.6)}{\pgfpoint{2.2cm}{1cm}}
\pgfputat{\pgfnodecenter{country}}{\pgfbox[center,center]{$\boldsymbol{\mu}_{\textrm{country}}$}}
\pgfputat{\pgfnodecenter{urban}}{\pgfbox[center,center]{$\boldsymbol{\theta}_{\textrm{urban}}$}}
\pgfputat{\pgfnodecenter{rural}}{\pgfbox[center,center]{$\boldsymbol{\theta}_{\textrm{rural}}$}}
\pgfputat{\pgfnodecenter{data1}}{\pgfbox[center,center]{Urban Data}}
\pgfputat{\pgfnodecenter{data2}}{\pgfbox[center,center]{Rural Data}}
\pgfsetendarrow{\pgfarrowto}
\pgfnodeconnline{urban}{data1}    \pgfnodeconnline{rural}{data2}
\pgfnodeconnline{country}{urban}   \pgfnodeconnline{country}{rural}
\end{pgfpicture}
\vspace{15mm}
\caption{A hierarchical structure for input parameters of EPP.}
\label{fig:hier}
\end{center}
\end{figure}

In the current EPP implementation, all areas apply the same prior in Equation (\ref{eqn:prior}) independently. We refer to the current model using prior (\ref{eqn:prior}) as the independent model. The availability and quality of the data vary from area to area. In areas with sufficient data, the independent model can work reasonably well, while in data scarce areas it would result in unreliable estimates. One way to overcome this issue in the data scarce areas is to borrow information from other areas with similar epidemic trends, such as areas within the same country. A hierarchical structure can be useful in associating parameters among areas, as illustrated in Figure \ref{fig:hier}. We see that the sub-national parameter estimates affect each other through the national parameter estimates.

For each country, let $\boldsymbol{\theta}_{i} = (t_{i0}, t_{i1},  r_{i0}, \beta_{i0}, \beta_{i1}, \beta_{i2}, \beta_{i3}, \beta_{i4})$ be the parameters of area $i$, so $\theta_{ij}$ is the $j$th parameter in the $i$th area, with $j=1,2,...,8$ and $i=1,2,...,K$. To implement a structure as in Figure \ref{fig:hier}, we can assume the following model:
\begin{eqnarray}
\theta_{ij} |\mu_j,\sigma_j^2 & \sim & N[\mu_j, (1-\rho_{0j}) \sigma_j^2], \nonumber\\
\mu_j|\sigma_j^2 & \sim & N[\mu_{0j},  \rho_{0j}\sigma_j^2 ], \label{eqn:hier}\\
\sigma_j^2 & \sim & \textup{InverseGamma}[a_{0j}, b_{0j}], \nonumber
\end{eqnarray}
where $\mu_{0j}$, $\rho_{0j}$, $a_{0j}$ and $b_{0j}$ are hyperparameters. Integrating out $\mu_j$ and $\sigma_j^2$, we get the joint distribution of $(\theta_{1j}, \theta_{2j}, \ldots, \theta_{Kj})$,
\begin{equation}
\begin{bmatrix}
\theta_{1j}\\
\theta_{2j}\\
\vdots\\
\theta_{Kj}
\end{bmatrix}
\sim \textup{MVT}_{(2a_{0j})}\left( 
\begin{bmatrix}
\mu_{0j}\\
\mu_{0j}\\
\vdots\\
\mu_{0j}
\end{bmatrix}, 
\sigma_{0j}^2 \times \begin{bmatrix}
1 & \rho_{0j} & \ldots & \rho_{0j}\\
\rho_{0j} & 1 & \ldots & \rho_{0j}\\
\ldots & \ldots & \ldots & \ldots\\
\rho_{0j} & \rho_{0j} & \ldots & 1
\end{bmatrix}
\right),
\label{eqn:cor}
\end{equation}
which is a multivariate $t$-distribution with $\sigma_{0j}^2=b_{0j}/a_{0j}$. Setting $a_{0j}=1$, $\mu_{0j}$ and $\sigma_{0j}$ to be the means and standard deviations in Equation (\ref{eqn:prior}), we have  the same marginal distributions between the hierarchical model (\ref{eqn:hier}) and the independent model (\ref{eqn:prior}) except for $t_0$. $t_0$ follows a uniform distribution ranging from 1970 to 1990 in the independent model, and a truncated t-distribution also ranging from 1970 and 1990 in the hierarchical model, and those two distributions have the same mean and variance. As a result, the hierarchical model and the independent model mainly differ in whether allowing correlations between parameters from different areas. 

As mentioned above, the independent prior (\ref{eqn:prior}) works reasonably well in areas with sufficient data. Moreover, even within the same country, different areas could have heterogeneous epidemics so that the parameters are not obviously correlated. While on the other hand, for some other areas with fewer data, the correlated structure (\ref{eqn:cor}) is critical in achieving a reliable estimate. In order for the priors to be flexible enough to accommodate both situations, we propose the following mixture model: let $f_{0j}$ be the prior density of the independent model (\ref{eqn:prior}) and $f_{1j}$ be the prior density of the hierarchical model (\ref{eqn:cor}) when $\rho_{0j} \ne 0$, and define the mixture model $g_j$ as the following:
\begin{equation}
g_j(\boldsymbol{\theta}_{j}) = \pi_{0j} f_{0j}(\boldsymbol{\theta}_{j}) + (1-\pi_{0j}) f_{1}(\boldsymbol{\theta}_{j}|\mu_{0j}, \sigma_{0j}^2, \rho_{0j}),
\label{eqn:mixture}
\end{equation}
where $\pi_{0j}$ is the mixing probability or the weight of the independent model that we will determine together with $\rho_{0j}$. The resulting mixture model (\ref{eqn:mixture}) accounts for the model uncertainty. 

\subsection{Hyperparameters}

We first need to determine a reasonable set of the hyperparameters $\boldsymbol{\rho_0}$ and $\boldsymbol{\pi_0}$. Here we adopt the empirical Bayes framework by using the point estimates from countries with multiple datasets. Let $\hat \theta_{icj}$ be the posterior mean of the $j$th parameter obtained from fitting the independent model to the $i$th area of a high-data-quality country $c$. $\pi_{0j}$ and $\rho_{0j}$ could be estimated by maximizing the likelihood with known $\mu_{0j}$ and $\sigma_{0j}^2$:
\begin{equation}
\prod_c g_j(\boldsymbol{\hat \theta}_{cj}) = \prod_c \left[ \pi_{0j} f_0(\boldsymbol{\hat \theta}_{cj}) + (1-\pi_{0j}) f_1(\boldsymbol{\hat \theta}_{cj}|\mu_{0j}, \sigma_{0j}^2, \rho_{0j}) \right]
\end{equation}

Currently, there are not many countries with high-quality datasets at sub-national level. \shortciteN{Fosdick2012} compared approaches of estimating the correlation in bivariate normal data with known variances and small sample sizes, and found that the Bayesian approach consistently outperformed several exact and approximate maximum likelihood estimators for small samples. Therefore, we assume a prior for $\rho_{0j}$ and $\pi_{0j}$ and set the posterior mode as the hyperparameter value in Equation (\ref{eqn:mixture}). Specifically, we assume that $\frac{\rho_{0j}+1}{2} \sim \textup{Beta}(1.5, 1.5)$ which is slightly concentrated towards $\rho_{0j}=0$ with density 1.273 at $\frac{\rho_{0j}+1}{2}=0.5$, while the uniform distribution has density 1 over the domain of $[0,1]$. We also assume that $\pi_{0j} \sim \textup{Beta}(1, 1.5)$ which has mean 0.6 and density 1.5 at 1 so that it is slightly in favor of larger values of $\pi_{0j}$. Finally, we denote $g=\prod_j g_j$ as the joint prior distribution of the mixture model and $f_0=\prod_j f_{0j}$ as the joint prior distribution of the independent model.
 
\subsection{Parameter Estimation via Importance Sampling}

Estimating parameters in dynamic models is often time consuming due to the lack of an analytic solution, the multi-modality and the nonlinearity. Fitting the datasets from multiple areas simultaneously in the mixture model (\ref{eqn:mixture}) will further increase the computing cost. It is desirable to develop a procedure that produces the mixture model results without additional computational cost. Since running the dynamic model is the most time consuming part of the estimation, we propose applying the independent model to fit each individual dataset as currently being implemented in EPP/Spectrum 2013 \shortcite{Brown2014} and using importance sampling to approximate the joint posterior distribution of parameters in the mixture model for multiple datasets. 

Applying the independent model to each individual dataset, the posterior samples are drawn by the incremental mixture importance sampling (IMIS) algorithm which has been proved to be efficient for estimating parameters in dynamic systems with a moderate number of parameters \shortcite{Raftery2010,Brown2010}. It works as follows:
\begin{enumerate}
\item Draw initial samples from a sampling distribution, e.g. the prior distribution, $f_0$;
\item Update the importance weight for each sample, which is the ratio between posterior density, $P(\boldsymbol{\theta_{i}}|\textrm{Data}_i)$, and density of the sampling distribution;
\item Find the sample with the highest weights, and draw new ones from a multivariate Gaussian distribution centered around it;
\item Combine new samples with all existing samples, update the sampling distribution by a mixture of initial sampling distribution and multiple Gaussian components, and recalculate the importance weight;
\item Iterate between step 2 and 4 until there is no large importance weight;
\item Resample all samples from the multinomial distribution with weights.
\end{enumerate}
The initial sampling distribution is often flat so that it ensures a good coverage of the entire parameter space. As a new Gaussian sampler is placed in the region that is under-represented by the sampling distribution at each iteration, the sampling distribution gets closer and closer to the posterior distribution.

To avoid a separate run of the proposed model, the step 6 of the IMIS algorithm can be revised as follows:
\begin{enumerate}
\item[6.a.] Repeat steps $1\sim5$ for areas $i=1,\ldots,K$. All samples with non-ignoble importance weight will be stored, e.g. greater than $10^{-6}$,;
\item[6.b.] Create a joint sample of parameters across all areas by randomly taking one sample from each area, and repeat until we have a large number of candidate joint samples, e.g. 1,000,000; 
\item[6.c.] The importance weight of a joint sample $(\boldsymbol{\theta_{1}}, \ldots, \boldsymbol{\theta_{K}})$ is calculated as the product of the importance weights for $\boldsymbol{\theta_{i}}$'s times $\frac{g(\boldsymbol{\theta_{1}}, \ldots, \boldsymbol{\theta_{K}})}{f_0(\boldsymbol{\theta_{1}}, \ldots, \boldsymbol{\theta_{K}})}$, the ratio between the mixture model prior density and the independent model prior density;
\item[6.d.] Resample all samples from the multinomial distribution with weights.
\end{enumerate}

The resulting posterior samples will have replicates because they are drawn from multinomial distribution with replacement, and a large number of replicates suggests a low sampling efficiency -- a large discrepancy between the sampling distribution and the target distribution \shortcite{Raftery2010}. The ratio between two prior densities is 
\[
\frac{g(\boldsymbol{\theta_{1}}, \ldots, \boldsymbol{\theta_{K}})}{f_0(\boldsymbol{\theta_{1}}, \ldots, \boldsymbol{\theta_{K}})} = \prod_j \left[ \pi_{0j}+(1-\pi_{0j}) \times \frac{f_{1j}(\theta_{1j}, \ldots, \theta_{Kj})}{f_{0j}(\theta_{1j}, \ldots, \theta_{Kj})} \right],
\]
which has the lower bound $\prod_j \pi_{0j}$, so that all combinations have non-ignoble reweighing factor. Therefore, limiting the importance weights helps obtain desired sampling efficiency. In the result section we will show that the revised IMIS algorithm still provides a large number of unique samples. 


\subsection{Assessing Model Fit}
We compare performance of the original EPP model (independent model) and the proposed mixture model based on datasets collected from the following 20 countries in Sub-Saharan Africa: Benin, Botswana, Burkina Faso, Burundi, Cameroun, Chad, Ethiopia, Ghana, Kenya, Lesotho, Malawi, Mali, Nigeria,  Republic of Cote d'Ivoire (RCI), Democratic Republic of the Congo (RDC), Rwanda, SierraLeone, Tanzania, Uganda, and Zambia. We construct two scenarios to assess the model fit by splitting the data into training and test in two different ways. In the first scenario we assess the prediction into the future by withholding the last 3 years of data as the test set. In the second scenario we assess both the future prediction and recovery of the past by using only the mid-period data as the training set and predict both the earlier and later data.

\section{Results}
\label{sect-Results}
To determine the values of $\boldsymbol{\pi_0}$ and $\boldsymbol{\rho_0}$, we obtain a set of posterior means by applying the independent model to 14 countries with multiple high-quality datasets, where high-quality is defined as having prevalence data at multiple clinics for at least 7 years and at least one national population based survey. Here we focus on high-quality datasets because their point estimates are mainly driven by the data instead of the informative prior distribution in Equation (\ref{eqn:prior}). Table \ref{tab:hyper} presents the estimated $\boldsymbol{\pi_0}$ and $\boldsymbol{\rho_0}$ following the procedure in Section 2.3. It suggests that the correlations of $\beta_2$'s between areas are negligible with probability $\pi_{06}=1$ of using the independent model prior distribution. 

\begin{table}[!h]
\caption{$\boldsymbol{\pi_0}$ and $\boldsymbol{\rho_0}$ determined by point estimates of high-quality datasets.}
\centering
\begin{tabular}{|l|rrrrrrrr|}
\hline
\multirow{2}{*}{ } & $\theta_1$ & $\theta_2$ & $\theta_3$ & $\theta_4$ & $\theta_5$ & $\theta_6$ & $\theta_7$ & $\theta_8$\\
& $t_0$ & $t_1$ & $\log r_0$ & $\beta_0$ & $\beta_1$ & $\beta_2$ & $\beta_3$ & $\beta_4$ \\ 
\hline
$\pi_{0j}$ & 0.278 & 0.249 & 0.315 & 0.154 & 0.490 & 1.000 & 0.196 & 0.125 \\ 
$\rho_{0j}$ & 0.822 & 0.925 & 0.996 & 0.708 & 0.783 & 0.000 & 0.529 & 0.608 \\ 
\hline
\end{tabular}
\label{tab:hyper}
\end{table}

We then apply both the independent model and the mixture model to datasets in 20 countries that have sufficient data to evaluate the prediction performance by with holding the last 3 years as test datasets. Table \ref{tab:mae1} summarizes prediction performance in term of the mean absolute errors (MAE) of antenatal clinic prevalence, and the sampling efficiency in term of the number of unique posterior samples. 

Despite the success of the independent model with the informative prior distribution in many high-quality-data countries \cite{Bao2012rtrend}, the mixture model further improves 3-year prediction MAE by 0.05 on average, with (5th, 95th) percentiles being (-0.03, 0.22). Among 41 datasets, the improvement is greater than 0.05 for 13 datasets, and less than -0.05 for one dataset. We also note that the mixture model provides better prediction accuracy especially for areas that have fewer data. Table \ref{tab:mae1} highlights eight areas with less than 5 years of training data -- four of them gain substantial improvements by using the mixture model, and the remaining four areas have little difference.

\begin{figure}[h]
\begin{tabular}{cc}
\begin{minipage}{6.5cm}
\includegraphics[width=6.5cm]{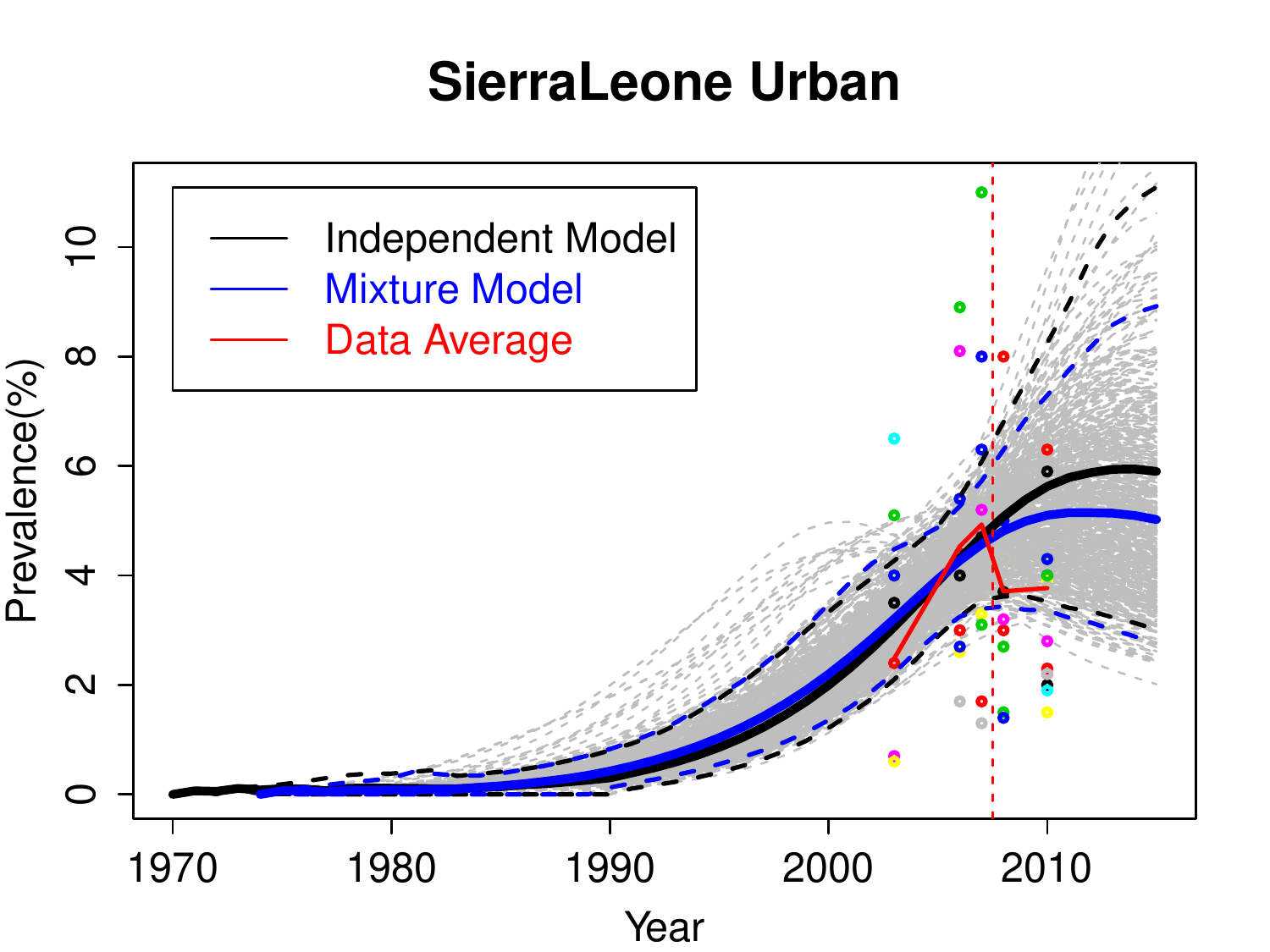}
\end{minipage}
&
\begin{minipage}{6.5cm}
\includegraphics[width=6.5cm]{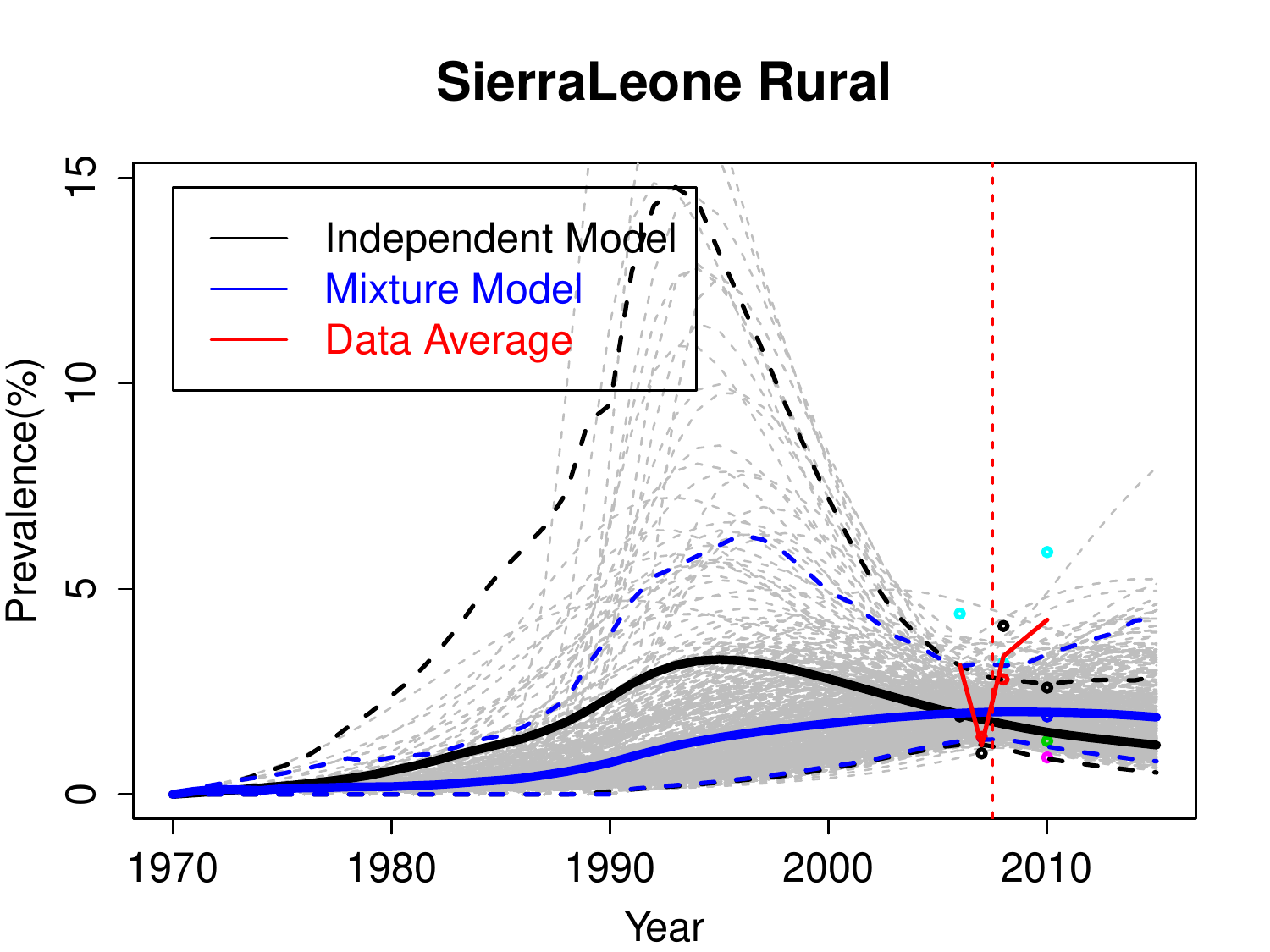}
\end{minipage}
\\
\begin{minipage}{6.5cm}
\includegraphics[width=6.5cm]{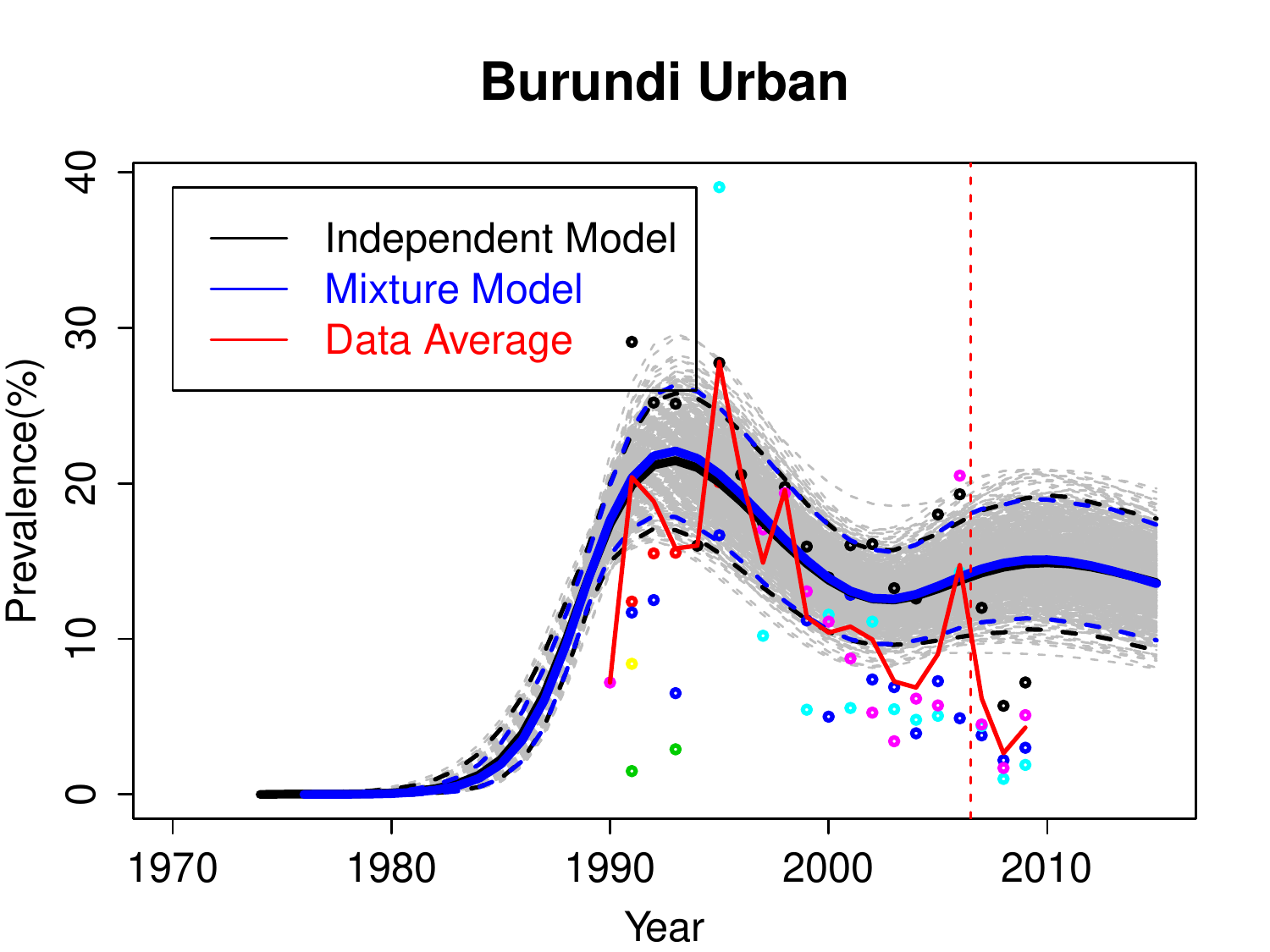}
\end{minipage}
&
\begin{minipage}{6.5cm}
\includegraphics[width=6.5cm]{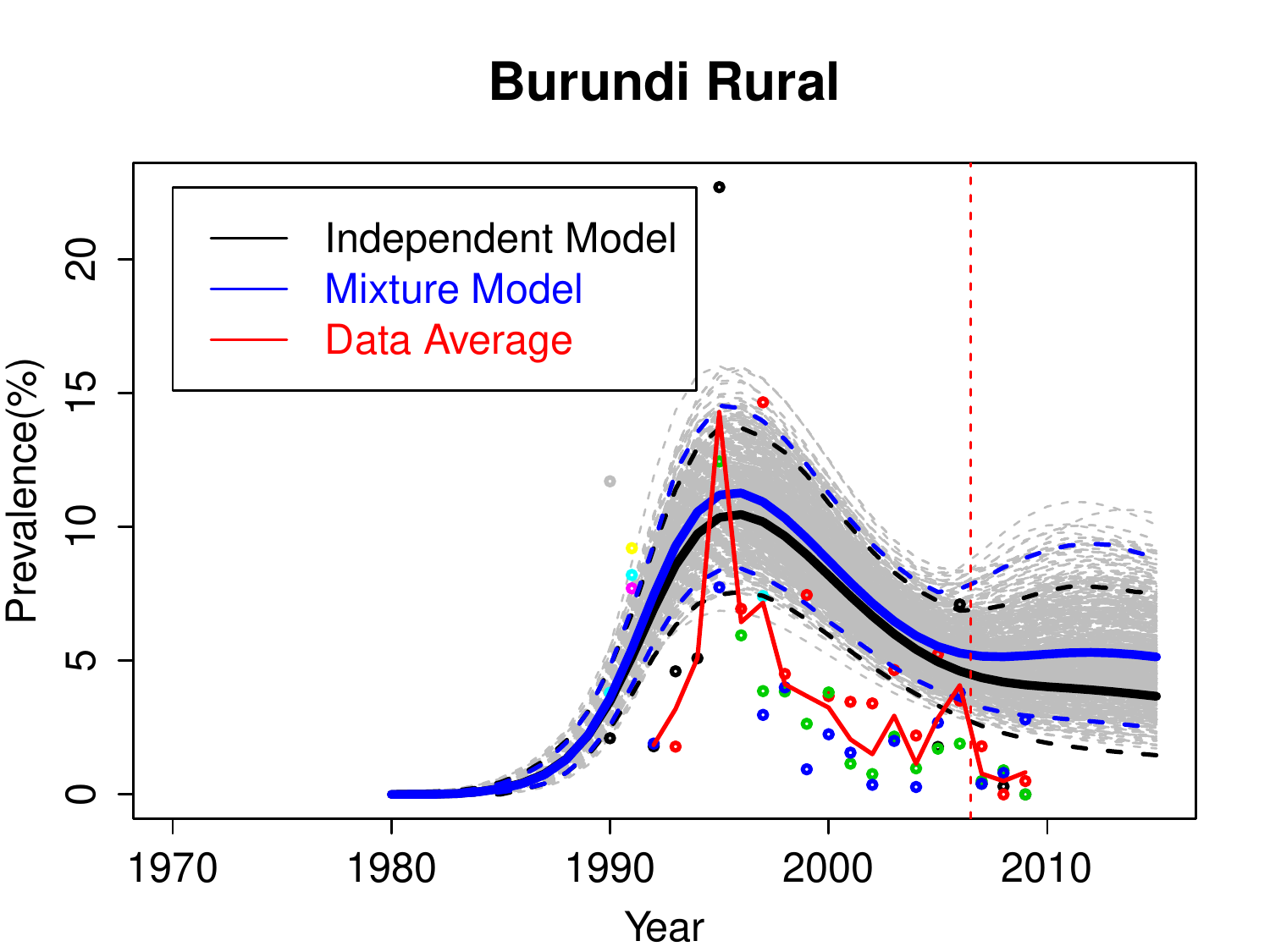}
\end{minipage}
\\
\begin{minipage}{6.5cm}
\includegraphics[width=6.5cm]{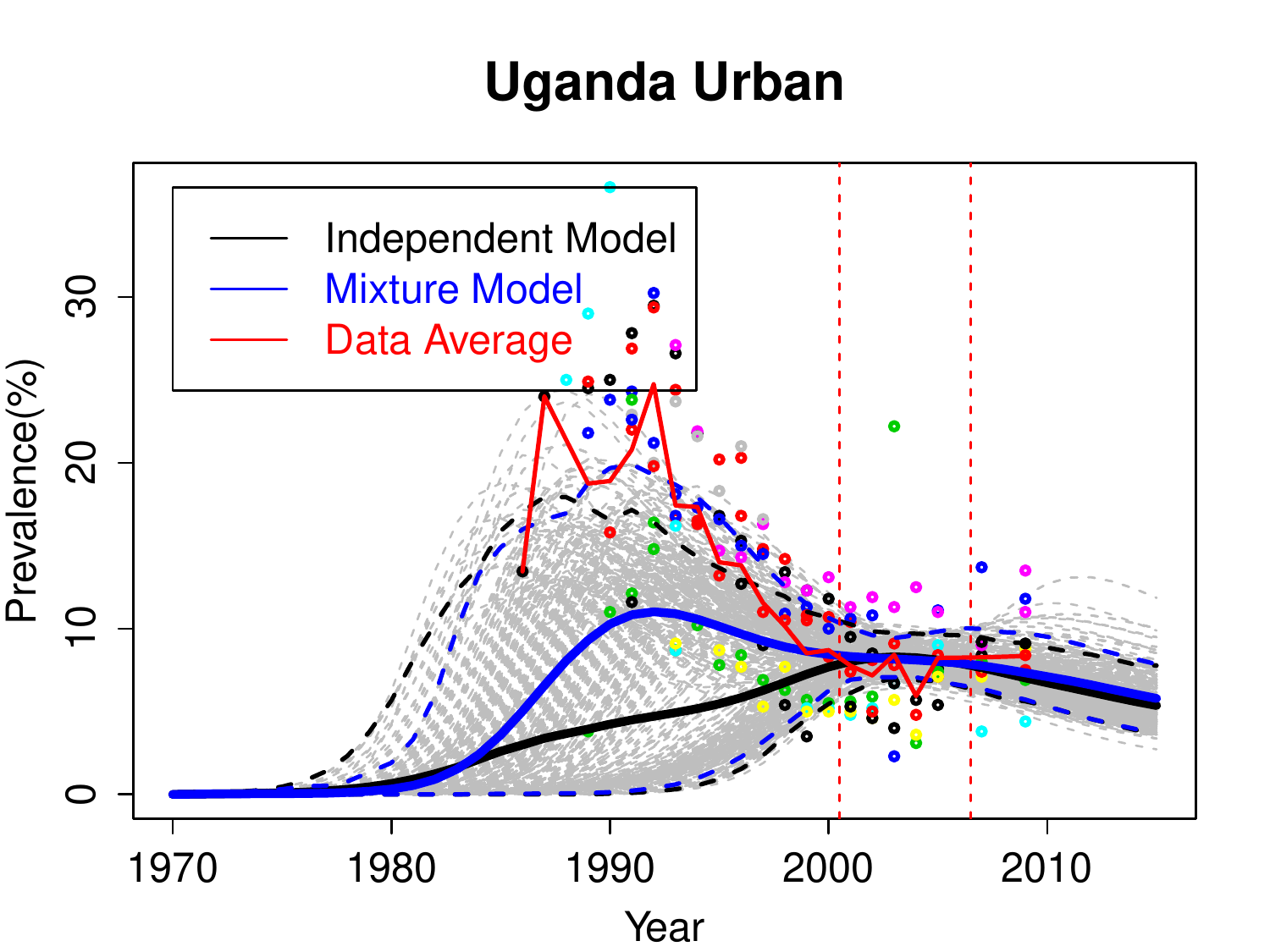}
\end{minipage}
&
\begin{minipage}{6.5cm}
\includegraphics[width=6.5cm]{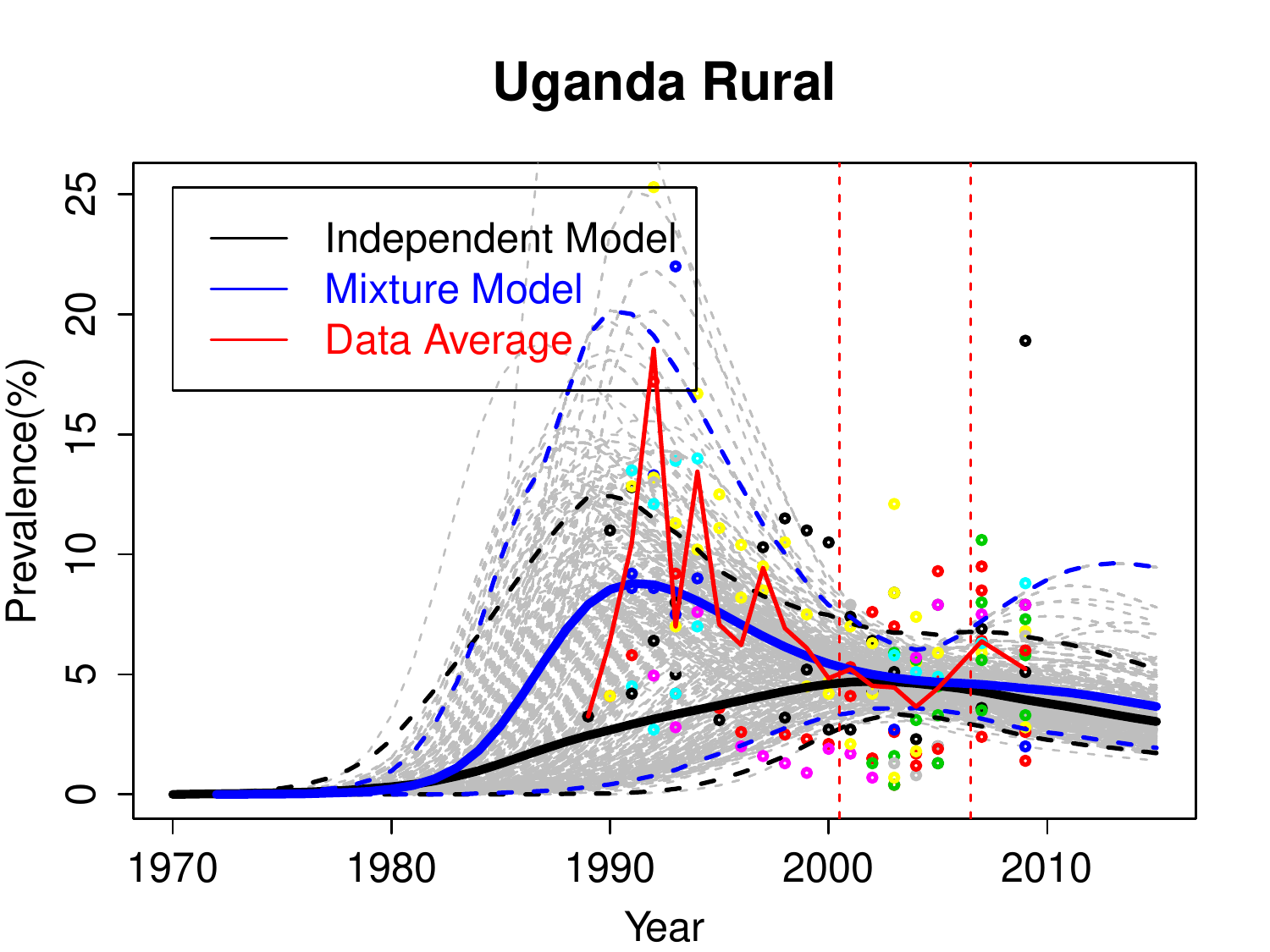}
\end{minipage}
\\
\end{tabular}
\caption{\footnotesize{HIV prevalence estimation and prediction for 6 datasets that present large differences between the independent model and the mixture model. Within each figure, the black solid line is the posterior mean trajectory of the original model; the  blue solid line is the posterior mean trajectory of the mixture model; the dashed black/blue lines show the 95\% credible intervals; colored dots are observed prevalence from different ANC sites; the red solid line is the raw average trajectory of ANC prevalence. The red dashed vertical lines separate the training data and test data.}}
\label{fig:prev_comp}
\end{figure}

The top four panels in Figure \ref{fig:prev_comp} present examples that the mixture model results differ systematically from the independent model results during the 3-year period of test data. In the Sierra Leone example, the mixture model (blue) improves the prediction accuracy as its posterior mean is closer to the average of observed HIV prevalence (red) than the posterior mean of independent model (black). Moreover, Sierra Leone surveillance data starts from 2003 in urban areas and 2006 in rural areas. Two models provided significant different estimates for the early period of Sierra Leone rural epidemic. It was because Sierra Leone urban data suggested the epidemic started after 1980 and the rural area estimation borrowed that information in the mixture model. In the Burundi example, the mixture model (blue) fails to make a better prediction. It was because the prevalence had an unexpected decline in the test data period (after the red vertical line), and neither urban training data nor rural training data could predict that decline. It revealed a limitation of the mixture model: if there is a sudden change of the epidemic pattern that could not predicted by any individual dataset, borrowing information from each other could not help in capturing the change.

IMIS method \cite{Raftery2010} is used to draw the posterior samples for the independent model. Among all datasets, it provides 801 unique samples out of 1,000 resamples on average, with (5th, 95th) percentiles being (767, 828). Adjusting the importance weights following procedures 6.a.$\sim$6.c. in section 2.4., the average number of unique samples for the mixture model is 537, and (5th, 95th) percentiles are (235, 762). It is less efficient than applying IMIS from the scratch in term of the number of unique samples. However, we avoid re-running the time-consuming EPP models and get new results within 10 seconds for each country. The gray curves in Figure \ref{fig:prev_comp} show the posterior samples of mixture model. It seems that simply adjusting the importance weights provides enough unique samples to cover the high posterior density region. 

\begin{table}[ht]
\caption{\footnotesize{A summary table for the number of data-years in training dataset, the 3-year prediction mean absolute error of antenatal clinic prevalence in \%, the number of unique posterior samples. Bold highlights the areas with less than 5 years of training data.}} 
\scriptsize
\centering
\begin{longtabu}{|ll|c|ccc|cc|}
  \hline
\multirow{2}{*}{Country} & \multirow{2}{*}{Area} & Training Data & \multicolumn{3}{c}{Mean Absolute Error (\%)} & \multicolumn{2}{|c|}{Number of Unique Samples} \\ 
					   &					    & $\#$ of Years & Independent & Mixture & Improvement & Independent & Mixture \\ 
  \hline
  Benin & Urban & 18 & 1.56 & 1.57 & -0.01 & 808 & 152 \\ 
  Benin & Rural & 6 & 0.88 & 0.88 & -0.01 & 824 & 186 \\ 
  Botswana & Urban & 15 & 3.07 & 3.07 & 0.01 & 657 & 460 \\ 
  Botswana & Rural & 15 & 4.31 & 4.14 & 0.17 & 793 & 467 \\ 
  BurkinaFaso & Urban & 17 & 0.71 & 0.7 & 0.02 & 797 & 513 \\ 
  BurkinaFaso & Rural & 5 & 0.47 & 0.48 & 0.00 & 821 & 360 \\ 
  Burundi & Urban & 17 & 6.66 & 6.69 & -0.03 & 814 & 730 \\ 
  Burundi & Rural & 17 & 1.52 & 1.74 & -0.22 & 818 & 571 \\ 
  Cameroun & Urban & 13 & 3.88 & 3.88 & 0.00 & 817 & 677 \\ 
    \rowfont{\bfseries}
  Cameroun & Rural & 2 & 3.3 & 3.03 & 0.26 & 803 & 613 \\ 
  Chad & Urban & 6 & 2.42 & 2.37 & 0.05 & 811 & 283 \\ 
    \rowfont{\bfseries}
  Chad & Rural & 2 & 1.23 & 1.01 & 0.22 & 809 & 244 \\ 
  Ethiopia & Urban & 11 & 2.5 & 2.43 & 0.08 & 785 & 283 \\ 
  Ethiopia & Rural & 7 & 1.22 & 1.22 & -0.01 & 808 & 374 \\ 
  Ghana & Urban & 14 & 0.74 & 0.72 & 0.02 & 803 & 450 \\ 
  Ghana & Rural & 13 & 0.71 & 0.75 & -0.04 & 822 & 430 \\ 
  Kenya & Urban & 15 & 3.11 & 3.04 & 0.07 & 759 & 624 \\ 
  Kenya & Rural & 12 & 2.72 & 2.5 & 0.22 & 819 & 617 \\ 
  Lesotho & Urban & 8 & 5.93 & 5.9 & 0.03 & 815 & 755 \\ 
  Lesotho & Rural & 8 & 4.17 & 4.01 & 0.16 & 817 & 772 \\ 
  Malawi & Central & 9 & 2.55 & 2.58 & -0.01 & 806 & 719 \\ 
  Malawi & Northern & 9 & 2.04 & 2.05 & -0.02 & 817 & 817 \\ 
  Malawi & Southern & 9 & 5.35 & 5.27 & 0.08 & 797 & 729 \\ 
    \rowfont{\bfseries}
  Mali & Urban & 3 & 1.03 & 1 & 0.03 & 803 & 665 \\ 
    \rowfont{\bfseries}
  Mali & Rural & 3 & 0.99 & 0.97 & 0.02 & 828 & 663 \\ 
  Nigeria & Urban & 7 & 2.58 & 2.59 & -0.01 & 782 & 725 \\ 
    \rowfont{\bfseries}
  Nigeria & Rural & 2 & 2.21 & 2.22 & -0.01 & 821 & 645 \\ 
  RCI & Urban & 14 & 1.97 & 1.92 & 0.05 & 795 & 528 \\ 
    \rowfont{\bfseries}
  RCI & Rural & 3 & 2.54 & 2.52 & 0.02 & 824 & 544 \\ 
  RDC & Urban & 20 & 1.23 & 1.22 & 0.02 & 770 & 670 \\ 
  RDC & Rural & 11 & 1.33 & 1.34 & -0.01 & 817 & 525 \\ 
  Rwanda & Urban & 6 & 1.88 & 1.8 & 0.09 & 807 & 529 \\ 
  Rwanda & Rural & 6 & 0.87 & 0.87 & 0.01 & 807 & 355 \\ 
    \rowfont{\bfseries}
  SierraLeone & Urban & 3 & 2.7 & 2.49 & 0.20 & 794 & 591 \\ 
    \rowfont{\bfseries}
  SierraLeone & Rural & 2 & 1.78 & 1.6 & 0.18 & 834 & 509 \\ 
  Tanzania & Urban & 18 & 1.61 & 1.61 & 0.00 & 796 & 685 \\ 
  Tanzania & Rural & 18 & 1.37 & 1.36 & 0.00 & 796 & 645 \\ 
  Uganda & Urban & 20 & 2.34 & 2.33 & 0.01 & 767 & 763 \\ 
  Uganda & Rural & 17 & 1.59 & 1.59 & 0.00 & 807 & 797 \\ 
  Zambia & Urban & 5 & 2.55 & 2.5 & 0.05 & 797 & 484 \\ 
  Zambia & Rural & 5 & 3.08 & 3.1 & -0.02 & 818 & 541 \\ 
   \hline
\multicolumn{3}{|l|}{Mean} 		& 2.23 & 2.18 & 0.05 & 801 & 537 \\
\multicolumn{3}{|l|}{5th quantile} 	& 0.71 & 0.72 & -0.03 & 767 & 235\\
\multicolumn{3}{|l|}{95th quantile}	& 4.55 & 4.40 & 0.22 & 828 & 762\\
  \hline
\addtocounter{table}{-1} 
\end{longtabu}
\label{tab:mae1}
\end{table}

In the next few years, many countries are going to produce estimates at finer sub-national scales, e.g. at province or district levels. The data imbalance across sub-national areas is a major concern -- some areas have historical surveillance sites with many years of observations, and some other area only have newly established surveillance sites with observations in recent years. One motivation of our model is to improve the estimation and prediction for low-data-quality areas by borrowing information from high-data-quality areas. Trying to mimic this situation, for each country, we create one ``low-data-quality'' area by keeping only the middle 6 years data as the training set and using both the early years data (however many years available) and last 3 years data as the test set in that area, and the other areas in that country act as ``high-data-quality'' areas such that only the last 3 years data are withheld as the test set. We also rotate the areas of ``low'' and ``high'' so each area gets evaluated as the ``low-data-quality'' area. We summarize the MAE's for both future predictions and past recoveries in Table \ref{tab:mae2}. The ``Area'' column indicates which area is the ``low-data-quality'' area. The results show that the mixture model improves 3-year prediction MAE by 0.05 on average with improvement greater than 0.10 in 8 cases and no less than -0.10 in all other cases. The mixture model also improves the early epidemic estimation MAE by 0.55 on average with improvement greater than 1.00 in 8 cases but less than -1.00 in one case. The bottom two panels in Figure \ref{fig:prev_comp} present two cases with large difference between two models in the early period of the epidemic, and the mixture model mean trajectory is a better approximation to the raw data average than the independent model.

\begin{table}[ht]
\caption{\footnotesize{A summary table for the early data period estimation mean absolute errors and the 3-year prediction mean absolute errors in \%.}}
\scriptsize
\centering
\begin{longtabu}{|ll|ccc|ccc|}
  \hline
\multirow{2}{*}{Country} 	& \multirow{2}{*}{Area} 		& \multicolumn{3}{c|}{Early Data Period MAE} 		& \multicolumn{3}{c|}{Prediction MAE} \\ 
					   		&					    		& Independent & Mixture & Improvement 					& Independent & Mixture & Improvement\\ 
  \hline
  Botswana & Urban & 9.44 & 3.60 & 5.84 & 3.17 & 2.98 & 0.18 \\ 
  Botswana & Rural & 4.6 & 6.16 & -1.57 & 4.39 & 4.33 & 0.06 \\ 
  Burundi & Urban & 7.79 & 7.84 & -0.05 & 16.57 & 16.65 & -0.08 \\ 
  Burundi & Rural & 3.83 & 3.94 & -0.11 & 2.15 & 2.25 & -0.10 \\ 
  Ethiopia & Urban & 3.40 & 3.20 & 0.20 & 2.29 & 2.16 & 0.14 \\ 
  Ethiopia & Rural & 1.44 & 1.45 & -0.01 & 1.16 & 1.16 & 0.00 \\ 
  Ghana & Urban & 1.89 & 1.24 & 0.65 & 0.84 & 0.81 & 0.03 \\ 
  Ghana & Rural & 1.50 & 0.91 & 0.59 & 0.71 & 0.76 & -0.05 \\ 
  Kenya & Urban & 5.74 & 5.69 & 0.05 & 1.87 & 1.84 & 0.03 \\ 
  Kenya & Rural & 4.41 & 4.31 & 0.10 & 2.52 & 2.54 & -0.02 \\ 
  Lesotho & Urban & 7.74 & 6.57 & 1.17 & 1.18 & 1.18 & 0.01 \\ 
  Lesotho & Rural & 10.03 & 8.23 & 1.80 & 2.97 & 2.77 & 0.20 \\ 
  Malawi & Central	& 3.82 & 4.77 & -0.85 & 2.20 & 2.25 & -0.05 \\ 
  Malawi & Northern 	& 7.63 & 7.18 & 0.45 & 1.72 & 1.71 & 0.01 \\ 
  Malawi & Southern 	& 5.67 & 3.64 & 2.03 & 4.65 & 4.74 & -0.09 \\ 
  RDC & Urban & 2.18 & 2.09 & 0.09 & 1.20 & 1.20 & 0.00 \\ 
  RDC & Rural & 3.27 & 2.46 & 1.19 & 0.85 & 0.85 & 0.00 \\ 
  Rwanda & Urban & 9.88 & 10.14 & -0.26 & 1.91 & 1.70 & 0.21 \\ 
  Rwanda & Rural & 1.51 & 1.84 & -0.33 & 1.13 & 0.93 & 0.20 \\ 
  Tanzania & Urban & 4.85 & 4.95 & -0.10 & 1.55 & 1.54 & 0.00 \\ 
  Tanzania & Rural & 7.88 & 6.53 & 1.35 & 1.61 & 1.35 & 0.27 \\ 
  Uganda & Urban & 8.47 & 6.04 & 2.42 & 2.60 & 2.45 & 0.15 \\ 
  Uganda & Rural & 4.04 & 2.56 & 1.47 & 2.18 & 2.07 & 0.11 \\ 
  Zambia & Urban & 7.12 & 4.78 & 2.34 & 2.34 & 2.41 & -0.07 \\ 
  Zambia & Rural & 2.99 & 2.74 & 0.25 & 2.75 & 2.85 & -0.10 \\ 
   \hline
\multicolumn{2}{|l|}{Mean} 			& 5.01 & 4.46 & 0.55 & 2.71& 2.66 & 0.05 \\
\multicolumn{2}{|l|}{5th quantile} 	& 1.45 & 1.27 & -1.43 & 0.84 & 0.82 & -0.09\\
\multicolumn{2}{|l|}{95th quantile}	& 9.81 & 8.17 & 2.41 & 4.33 & 4.27 & 0.21\\
  \hline
\end{longtabu}
\label{tab:mae2}
\end{table}

\begin{figure}[h]
\begin{tabular}{cc}
\begin{minipage}{6.5cm}
\includegraphics[width=6.5cm]{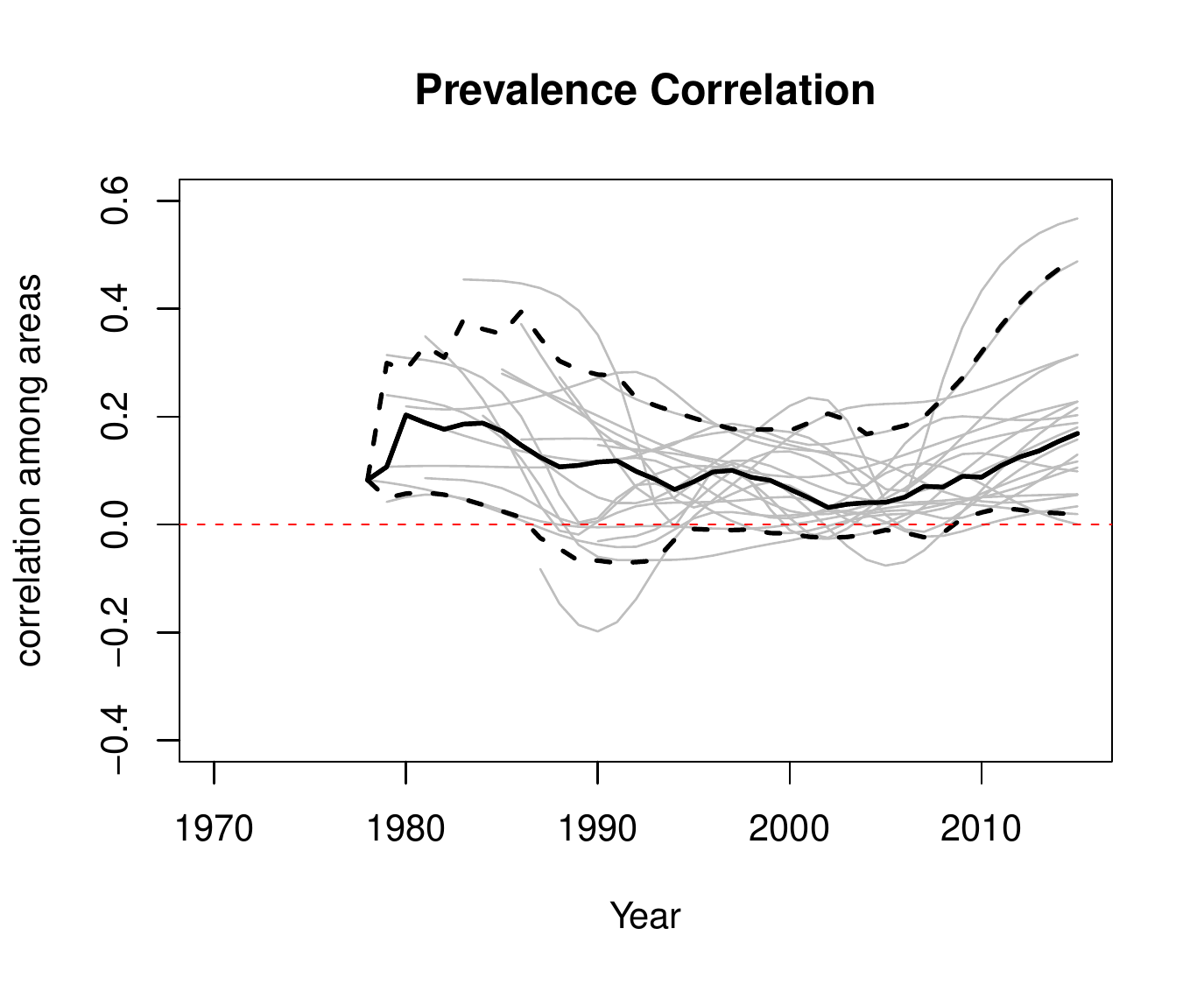}
\end{minipage}
&
\begin{minipage}{6.5cm}
\includegraphics[width=6.5cm]{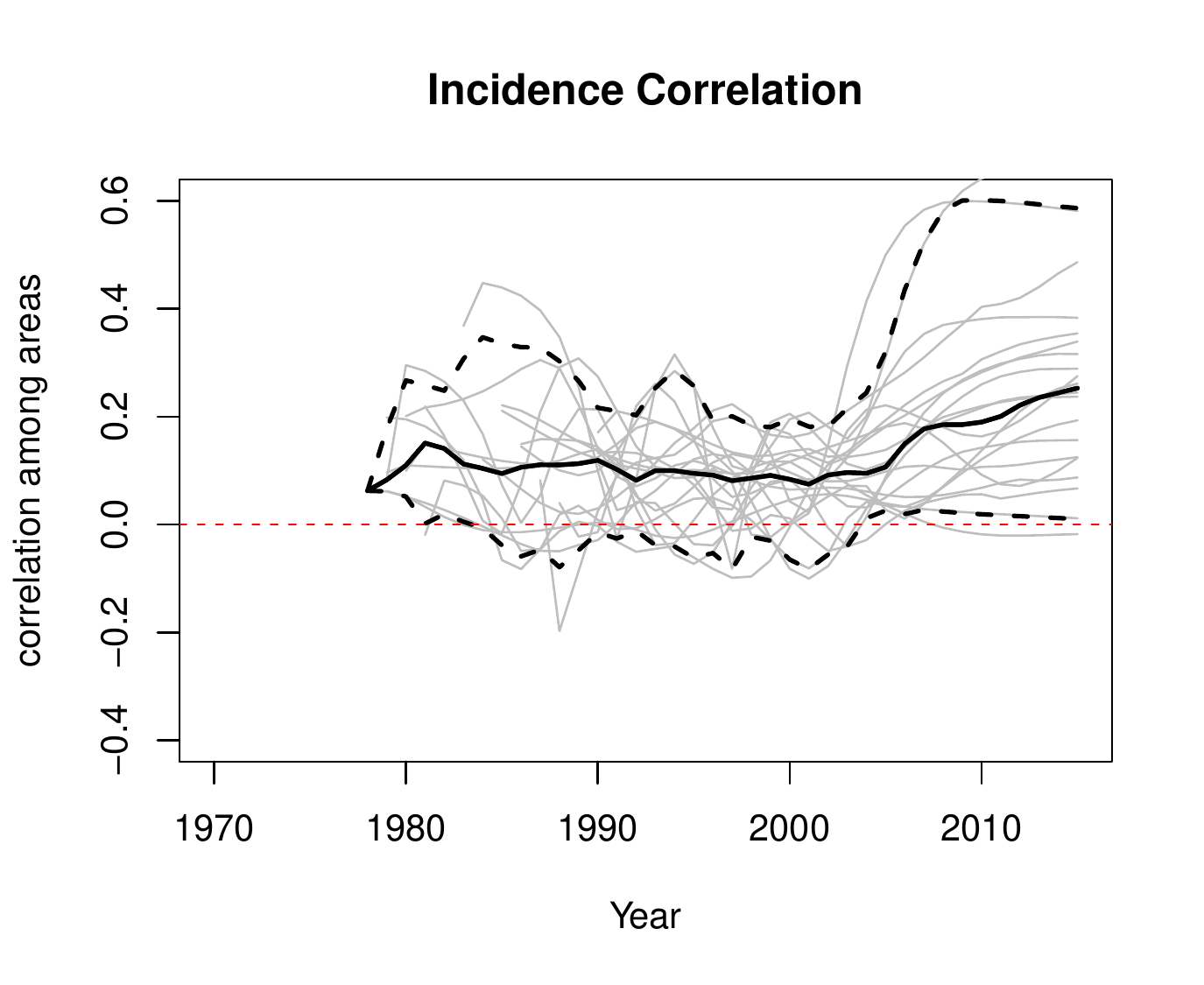}
\end{minipage}
\\
\end{tabular}
\caption{\footnotesize{The prevalence correlations (left) and incidece correlations (right) between subnaitonal areas produced by the mixture model. Each gray line corresponds to one country. The black solid line and dashed lines provide the median and (5th, 95th) quantiles. The red dashed line indicates a zero correlation which corresponds to the expected correlation in the independent model.}}
\label{fig:cor}
\end{figure}

In addition to the examination of the marginal distribution of prevalence within each area, we also examine the joint distributions of the quantity of interest across areas, e.g. HIV prevalence and HIV incidence. We apply the mixture model to the full datasets of 20 countries. For every country and every year, we calculate the prevalence correlation and incidence correlation among areas. Figure \ref{fig:cor} shows those correlations by years and by countries. Those correlations are mostly positive due to the positive correlation of EPP input parameters. Since the mixture model involves a hierarchical structure for the EPP input parameters which summarize the characteristics of the HIV epidemic, the correlations of the EPP output prevalence and incidence vary across time.

\section{Discussion}
\label{sect-Discussion}
Planning an effective response to the HIV epidemic and assessing the impact of the past response requires quantitative analysis. Sub-national estimates can be used for better program planning and management, such as assessing and meeting the needs for commodities, human resources and other program elements, measuring population coverage of treatments, and monitoring and evaluating interventions. In this article, we describe a hierarchical epidemiological structure that assumes the patterns of the sub-national epidemic are similar, and suggest a mixture between the independent model and newly proposed hierarchical model to account for the uncertainty between two model choices. For the estimation and projection of HIV epidemics, this will be the first attempt of developing a joint model for multiple areas within a country. It will be potentially implemented in the next round of EPP/Spectrum software (available from http://www.futuresinstitute.org). The proposed model can be easily generalized to the study of other epidemics. 

We find that the predictive ability of EPP model can be improved by using the mixture model, especially for the areas that are lack of long historical data. The mixture model also provides more accurate assessment of the early period of the epidemics in many examples, but there are a couple of exceptions. If the mixture model and the independent model estimate the early trend of the epidemic very differently, we suggest choosing the model with the help of other available information such as the year of the first case report in that area. The prediction accuracy might be further improved by redefining the default prior distribution of independent model, such as allowing dependence among $(t_0, t_1, r_0, \beta_0, \beta_1, \beta_2, \beta_3, \beta_4)$. However, it is beyond the focus of this article. 

The computing cost of estimating parameters in a dynamic model is often high due to the lack of an analytic solution, the multi-modality and nonlinearity. We propose an importance sampling method that draws posterior samples of parameters in the mixture model without needing to refit the data or unpack the existing software for implementing an independent model. We first obtain posterior samples of parameters for each datasets independently. Since the mixture model and the independent model only differ at their prior distributions, we can randomly combine posterior samples of the independent model from multiple areas, reweight the joint samples by the ratio between the mixture model prior and the default independent prior, and draw a set of joint posterior samples from the multinomial distribution with weights. The mixture between the original model prior and the new model prior provided a lower bound on the ratio, and thus provided satisfactory sampling efficiency as measured by the number of unique posterior samples in our examples. If the number of unique posterior samples is too small, we suggest either increasing the number of random combinations, or using a less informative prior distribution in the independent model. A more efficient sampling strategy should be developed in the future.

Note that, in this specific application, the incremental mixture importance sampling method is used to draw posterior samples for each individual data, but our proposed reweighting approach is generic and does not depend on how the posterior samples are drawn for individual datasets. For instance, if Markov chain Monte Carlo (MCMC) is used to draw posterior samples for individual datasets with reasonable burn-in and thinning, we can assume that those posterior samples have equal weights in the independent model. This suggests an alternative way of estimating parameters in dependent data, e.g. data with hierarchical/spatial/temporal structure: fitting each piece of data independently, and merge the results with adjustments for the difference between the independent model and the joint model. 

When a country has a large geographical territory or diverse regions of public health conditions, it is worthy to further develop a spatial model or multilevel hierarchical model based on the knowledge of the national officials. In those cases, the similarity of epidemic patterns may depend on the distance or adjacency status between areas, or the regions to which the areas belong. In countries with low-level and concentrated epidemics, HIV has spread rapidly in several high risk groups, but is not well established in the general population. Fewer data are available for those high risk populations due to the stigmatized nature of these populations in many countries. In such cases, a hierarchical model can allow sharing information across areas and high risk groups.

The results presented in this paper are based on illustrative HIV prevalence data for these countries, which may not be complete. These results should therefore not be seen as replacing or competing with official estimates regularly published by countries and UNAIDS.

\bibliographystyle{chicago}
\bibliography{EPP}


\end{document}